\documentclass{osa-article}

\journal{oe}

\usepackage{amsmath}

\usepackage{graphicx,color}           

\newcommand{\ex}{\mathbf{e}_x}
\newcommand{\ey}{\mathbf{e}_y}
\newcommand{\ez}{\mathbf{e}_z}

\DeclareMathAlphabet{\mathpzc}{OT1}{pzc}{m}{it}

\begin{document}

\title{Multiple-camera defocus imaging of ultracold atomic gases}

\author{
A.~R.~Perry,\authormark{1,2}
S.~Sugawa,\authormark{1,3,4} 
F.~Salces-Carcoba,\authormark{1,5}
Y.~Yue,\authormark{1}
and I.~B.~Spielman\authormark{1,*}}

\address{
\authormark{1}Joint Quantum Institute, National Institute of Standards and Technology, and University of Maryland, Gaithersburg, Maryland, 20899, USA\\
\authormark{2}Honeywell Quantum Solutions, 303 S. Technology Ct. 80021 Broomfield, Colorado, USA\\
\authormark{3}Institute for Molecular Science, National Institutes of Natural Sciences, Myodaiji, Okazaki 444-8585, Japan
\authormark{4}SOKENDAI (The Graduate University for Advanced Studies), Myodaiji, Okazaki 444-8585, Japan
\authormark{5}LIGO Laboratory, California Institute of Technology, MS 100--36,
Pasadena, CA 91125, USA
}

\email{\authormark{*}ian.spielman@nist.gov} 

\homepage{http://ultracold.jqi.umd.edu} 


\begin{abstract}
In cold atom experiments, each image of light refracted and absorbed by an atomic ensemble carries a remarkable amount of information.
Numerous imaging techniques including absorption, fluorescence, and phase-contrast are commonly used.
Other techniques such as off-resonance defocused imaging (ORDI,~\cite{Turner:05,Turner:04,Turner:Thesis,Wigley2016a}), where an in-focus image is deconvolved from a defocused image, have been demonstrated but find only niche applications.
The ORDI inversion process introduces systematic artifacts because it relies on regularization to account for missing information at some spatial frequencies. 
In the present work, we extend ORDI to use multiple cameras simultaneously at degrees of defocus, eliminating the need for regularization and its attendant artifacts.
We demonstrate this technique by imaging Bose-Einstein condensates, and show that the statistical uncertainties in the measured column density using the multiple-camera off-resonance defocused (MORD) imaging method are competitive with absorption imaging near resonance and phase contrast imaging far from resonance.
Experimentally, the MORD method may be incorporated into existing set-ups with minimal additional equipment.
\end{abstract}

\section{Introduction} \label{sec:Introduction}

Ultracold atoms exist in isolation, enshrouded in ultrahigh vacuum, so that nearly every measurement on them relies on their interaction with electromagnetic fields. The most common measurements use a probe laser beam that is attenuated and phase shifted by the atoms to recover two-dimensional images of the integrated density--the column density--of the atoms. Whether the technique be absorption imaging (AI), or phase-contrast imaging (PCI), the spatially resolved column density of the atomic cloud is recovered; from this, physical information regarding the atomic ensemble can be extracted.

\begin{figure}[t!]
\begin{center}
\includegraphics{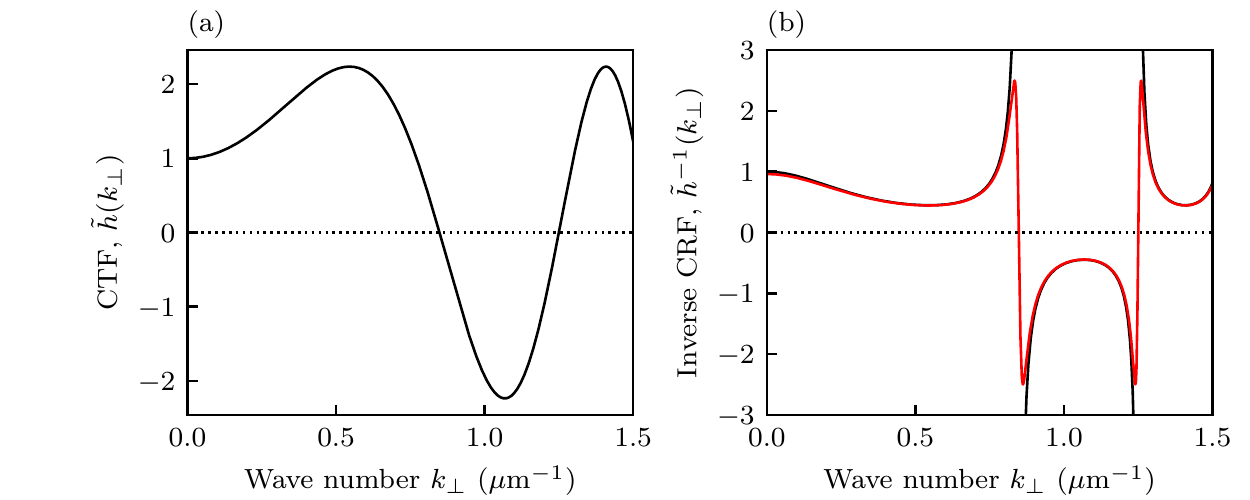}
\end{center}
\caption[Contrast transfer function]{Contrast transfer function (a) and its inverse (b), as discussed in Sect.~\ref{Section:OneCamera}, using eqns.~\eqref{eq:CTF} and \eqref{eq:CTFinv}.
These are computed as a function of transverse wave number $k_\perp$, for a probe laser beam with wavelength $\lambda=780\ {\rm nm}$, detuning $\bar\delta = 1$, and defocus distance $z = 60\ \mu{\rm m}$.
The black curves denote $\tilde h(k_\perp)$ and $\tilde h^{-1}(k_\perp)$ where appearance of repeating zeros in the CTF and the associated divergences in its inverse indicate spatial frequencies where all information about the initial density profile is lost.
The red curve shows the regularized inverse  $\tilde h^{-1}_{\rm R}(k_\perp)$ with regularization parameter $\eta = 0.2$, which tracks  $\tilde h^{-1}(k_\perp)$ until it exceeds a threshold value and then returns to zero.}
\label{CTFOneFullFig}
\end{figure}

In this paper, we describe and demonstrate an extension to off-resonance defocused imaging (ORDI) pioneered in Refs.~\cite{Turner:05,Turner:04,Turner:Thesis,Wigley2016a}.
ORDI uses information from both the absorbed and phase-shifted probe laser light; by contrast, absorption or phase-contrast imaging rely only on the absorption or phase shift signal, respectively.
Both the absorption and phase-shift are proportional to the quantity of interest, the column density.
When the laser detuning from atomic resonance $\delta$ is large, the fractional absorption $\propto 1/\delta^{2}$, while the phase-shift $\propto 1/\delta$: thus for sufficiently large detunings the phase-shift is more significant than absorption.
Typically AI~\cite{Ketterle:ImgNotes} is used for clouds of low to medium column density using a resonant probe beam (no phase-shift), and PCI~\cite{Ketterle:97} is used to image clouds of high column density using a far-detuned probe beam (negligible absorption). In AI and PCI, intensity images are recorded by a detector at the focus of the imaging system.
In ORDI, a image is taken by a detector positioned away from the focus: remarkably, Ref.~\cite{Turner:04} showed that given full knowledge of the atoms' complex susceptibility it is possible to digitally refocus intensity images of atoms by inverting contrast transfer function [CTF, shown in Fig.~\ref{CTFOneFullFig}(a)] without knowing the phase of the underlying optical field.
Still, ORDI was beset with unavoidable imaging artifacts resulting from lost information at some spatial frequencies in defocused images where the inverse CTF diverges [Fig.~\ref{CTFOneFullFig}(b)].
Here we demonstrate a technique to reconstruct defocuse images of ultracold atoms with greatly reduced artifacts.
In this technique, multiple-camera off-resonance defocused (MORD) imaging, we simultaneously use  cameras placed at different defocused distances and show that suitably placed cameras allow for essentially artifact-free reconstruction of the atomic column density.
We compare this technique to conventional imaging techniques and show that its signal to noise ratio (SNR) is comparable to AI near atomic resonance and comparable to PCI far from resonance.

This paper is organized as follows; in Sec.~\ref{sec:Introduction} we discuss the solution to the vector wave equation under the paraxial approximation after interacting with a thin, dilute atomic cloud.
In Sec.~\ref{sec:ImagingTechniques}, we describe absorption and phase-contrast imaging; derive expressions for ORDI and MORD imaging; and conclude with a theoretical comparison of MORD to absorption and phase-contrast imaging.
In Sec.~\ref{Sec:Techniques}, we describe the experimental implementation of the MORD method with three detectors, and the procedure we used to prepare $^{87}$Rb Bose-Einstein condensates (BECs) of variable column density.
In Sec.~\ref{sec:MeasAnalysis}, we present our experimental MORD results.

\subsection{Propagation of an electromagnetic wave} \label{sec:HelmholtzEqn}
First we introduce the theoretical problem, starting with the propagation of an electromagnetic wave defined at all positions ${\bf r} = x\ex + y\ey + z\ez$. 
For our neutral atomic systems interrogated by a monochromatic probe laser beam with wavelength $\lambda$ and wave number $k_{0}$\,=\,$2\pi / \lambda$, the evolution of the laser beam's electric field in the presence of atoms with complex susceptibility $\chi$ is described by a pair of scalar wave equations
\begin{equation}
\nabla^{2} \mathcal{E}_{i}({\bf r}) + k_{0}^{2} [1 + \chi({\bf r})] \mathcal{E}_{i}({\bf r}) = 0, \label{HelmholtzEqn}
\end{equation}
i.e., Helmholtz equations, one for each polarization component $\mathcal{E}_i$.
These equations are valid provided that the susceptibility $\chi$ changes slowly with respect to $\lambda$ [allowing the term $\nabla (\nabla \cdot \boldsymbol{\mathcal{E}})$ in the vector wave equation to be neglected].  
For each polarization component, the Helmholtz equation then has the formal solution
\begin{equation}
\mathcal{E}({\bf r}_\perp, z + \Delta z) = \exp \left\{ \pm i \Delta z \left[\nabla_{\bot}^{2} + k_{0}^{2} + \chi({\bf r}_\perp) k_{0}^{2} \right]^{1/2} \right\} \mathcal{E}({\bf r}_\perp, z), \label{WaveMedium}
\end{equation}
provided $\chi({\bf r})$ has no dependence on $z$.
Here ${\bf r}_\perp = x\ex + y\ey$ and $\nabla_{\bot}^{2}$\,=\,$\partial^{2}/\partial^{2}x + \partial^{2}/\partial^{2}y$ are the transverse position and Laplacian, respectively.
$\mathcal{E}({\bf r}_\perp, z)$ is the field before propagation, and $\mathcal{E}({\bf r}_\perp, z + \Delta z)$ is the field that has propagated an infinitesimal distance $\Delta z$ through the medium.
This solution becomes straightforward to evaluate numerically even for arbitrary displacements provided that $\chi$ has no spatial dependence -- such as the case for propagation in a homogenous isotropic medium -- by working in the Fourier domain.

In our simulations, we only used eq.~\eqref{WaveMedium} to evolve the electric field in free space where $\chi=0$.
For the problem at hand, we need to understand light traveling though a medium with $\chi({\bf r}_\perp, z)$ dependent on both ${\bf r}_\perp$ and $z$, a situation where eq.~\eqref{HelmholtzEqn} is inconvenient to manipulate analytically.
For these cases, we make the paraxial approximation that renders both numerical and analytical treatments straightforward. 
We write the field $\mathcal{E}({\bf r}_\perp,z)=\exp(i z k_{0}) E({\bf r}_\perp,z)$ making explicit the nominal propagation axis $\ez$.
In addition, we assume that the transverse spatial structure is slowly varying compared to the optical wavelength, i.e., $\left|\partial E/\partial (k_0 z) \right| \gg \left|\partial^2 E/\partial^2 (k_0 z) \right|$.
Together these approximations lead to the paraxial wave equation
\begin{align}
i \frac{\partial}{\partial z} E({\bf r}_\perp,z) &= \left[-\frac{\nabla_\perp^2}{2 k_0}+ \frac{k_0 \chi({\bf r}_\perp,z)}{2}\right]E({\bf r}_\perp,z).\label{eq:ParaxialWave}
\end{align}
In this manuscript, the paraxial wave equation offers two primary benefits.  Firstly the free space propagator
\begin{align*}
h_{\rm p}({\bf r}_\perp,\Delta z) &= \exp\left(i\Delta z\frac{\nabla_\perp^2}{2 k_0}\right), & \text{that gives} && E({\bf r}_\perp,z + \Delta z) = h_{\rm p}({\bf r}_\perp,\Delta z) E({\bf r}_\perp, z),
\end{align*}
is Gaussian and is therefore straightforward to manipulate analytically.  Secondly, we parametrize propagation through a medium localized at $z=0$ in terms of an absorption coefficient $\alpha ({\bf r}_\perp)$ and a phase shift $\phi ({\bf r}_\perp)$.  We relate the electric field $E_-({\bf r}_\perp, z=0)$ just prior to the medium to the field $E_+({\bf r}_\perp, z=0)$ just after the medium with
\begin{align}
E_+({\bf r}_\perp, z=0) &= \exp\left[i\frac{k_0}{2}\int \chi({\bf r}_\perp, \Delta z) dz \right]E_-({\bf r}_\perp, z=0)\nonumber\\
&= \exp\left[-\alpha({\bf r}_\perp) + i\phi({\bf r}_\perp)\right] E_-({\bf r}_\perp, z=0).\label{eq:ORDwave}
\end{align}
This expression is valid provided the medium is thin compared to the depth of field $2 k_0/k_{\rm max}^2$, where $k_{\rm max}$ is the largest wave-vector of any significance present in $E({\bf r})$, or more pragmatically $k_{\rm max}$ might be the largest wave-vector resolvable by the experimental detector.  
In effect, eq. \eqref{eq:ORDwave} neglects diffraction as light propagates through the atomic layer, and is equivalent to dropping the Laplacian in eq.~\eqref{eq:ParaxialWave}.

For two-level atoms, we characterize the absorption and phase shift via the complex susceptibility
\begin{equation}
\chi({\bf r}) = \frac{\sigma_{0}}{k_{0}}\left[ \frac{i-2 \bar\delta}{1+\bar I({\bf r})+4 \bar\delta^2} \right] \rho ({\bf r}),\label{Susceptibility}
\end{equation}
for an ensemble of atoms with density $\rho ({\bf r})$ illuminated by a probe laser of wavelength $\lambda$. Eq.~\eqref{Susceptibility} is valid for dilute clouds $\rho ({\bf r}) \ll k_{0}^{3}$ such that collective effects can be neglected.  Here $\sigma_{0}=3 \lambda^{2} / 2 \pi$ is the resonant cross-section, $\bar I({\bf r}) = I({\bf r})/I_{{\rm sat}}$ is the optical intensity $I({\bf r}) = c\epsilon_0\left|E({\bf r})\right|^2/2$ in units of the saturation intensity $I_{{\rm sat}}$, and $\bar\delta = \delta / \Gamma$ is the detuning $\delta$ from atomic resonance in units of the natural line-width $\Gamma$.  Here, $c$ is the speed of light and $\epsilon_0$, once known as the permittivity of free space, is the now ill-named electric constant.

The two-level model for the susceptibility gives coefficients
\begin{align}
\alpha({\bf r}_\perp) &= {\rm OD}({\bf r}_\perp)/2, & {\rm and} && \phi({\bf r}_\perp) &= -\bar\delta\ {\rm OD}({\bf r}_\perp).\label{eq:abs_phase}
\end{align}
Where the optical depth 
\begin{align}
{\rm OD}({\bf r}_\perp) &\equiv -\ln\left[\frac{I_+({\bf r}_\perp)}{I_-({\bf r}_\perp)}\right] = 2\alpha,\label{eq:OD_Def}
\end{align}
expresses exponential attenuation of light by the atoms and is defined in terms of the intensity just before $I_-({\bf r}_\perp)$ and just after $I_+({\bf r}_\perp)$ interacting with the atoms.  The column density
\begin{align*}
\rho_{\rm 2D}({\bf r}_\perp) &= \int \rho({\bf r}_\perp, z){\rm d}z
\end{align*}
can be derived from the optical depth given both the detuning and the intensity using~\cite{Reinaudi:07}
\begin{align}
\sigma_0 \rho_{\rm 2D}({\bf r}_\perp) &= (1 + 4\bar\delta^2) {\rm OD}({\bf r}_\perp) + \bar I_-({\bf r}_\perp)\left[1 - e^{-{\rm OD}({\bf r}_\perp)}\right].\label{eq:column_density}
\end{align}
This leads us to two essential messages for this section: (I) according to eq.~\eqref{eq:abs_phase} the complete impact of the atomic ensemble on the electric field can be parametrized in terms of ${\rm OD}({\bf r}_\perp)$ alone; and (II) once ${\rm OD}({\bf r}_\perp)$ is obtained, the column density can be reconstructed using eq.~\eqref{eq:column_density}, independent of what measurement technique was employed to obtain the optical depth. 

As a result, any imaging technique will first find the optical density and then obtain the column density using eq.~\eqref{eq:column_density}.  Therefore in the following discussion we will compare imaging techniques in terms of their ability to reconstruct the optical depth.

\section{Imaging techniques} \label{sec:ImagingTechniques}
In our experiments, the detectors -- CCD cameras -- are sensitive to the field's intensity, not its amplitude.  The imaging techniques discussed here use image pairs, each consisting of a 2D array of pixels giving the optical intensity with atoms present $\bar I^{(i)}_+({\bf r}_\perp)$ resulting from the field $E_+({\bf r}_\perp,z^{(i)})$, and the intensity with the atoms absent $\bar I^{(i)}_-({\bf r}_\perp)$ resulting from the field $E_-({\bf r}_\perp,z^{(i)})$.  The superscript $(i)$ indicates that for MORD several cameras simultaneously recordthe intensity at different positions $z^{(i)}$.   We characterize the impact of the atoms in terms of the fractional drop in intensity
\begin{align}
g^{(i)}({\bf r}_\perp) &= 1-\frac{I^{(i)}_+({\bf r}_\perp)}{I^{(i)}_-({\bf r}_\perp)}.\label{eq:FracIntensity}
\end{align}
Where it is clear, we will omit the camera index $i$ and pixel coordinate ${\bf r}_\perp$.  Evidently this detection process has no direct sensitivity to the fields' phase, nor the phase shift imprinted by the atoms. 

Each acquired image has a noise contribution $\delta I^{(i)}({\bf r}_\perp)$ that we model as a classical random process from shot noise on the CCD detector with
\begin{align}
\langle\delta I^{(i)}({\bf r}_\perp)\rangle &= 0, & {\rm and} && \langle\delta I^{(i)}({\bf r}_\perp)\delta I^{(i^\prime)}({\bf r}_\perp^\prime)\rangle &= \delta_{i,i^\prime}\delta_{{\bf r}_\perp,{\bf r}_\perp^\prime} I_0 \langle I^{(i)}({\bf r}_\perp) \rangle,
\end{align}
where $I_0 = {\hbar\omega}/{\eta A \Delta t}$ is the intensity required to generate a single photo-electron.  Here $\hbar\omega = c \hbar k_0$ is the single-photon energy, $\eta$ is the detector quantum efficiency (shot noise is really the shot noise of the photo-electrons not of the photons), $A$ is the area of a single pixel, and $\Delta t$ is the measurement time.  This noise model neglects other physical sources of noise such as CCD readout noise, or excess noise from dark current.

Since the fractional absorption contains the full information about the atomic ensemble, we transfer the noise in the images to the noise in the fractional intensity which has $\langle\delta g^{(i)}({\bf r}_\perp)\rangle = 0$ and the correlation function
\begin{align}
\langle\delta g^{(i)}({\bf r}_\perp)\delta g^{(i^\prime)}({\bf r}_\perp^\prime)\rangle &= \delta_{i,i^\prime}\delta_{{\bf r}_\perp,{\bf r}_\perp^\prime} \frac{I_0}{I^{(i)}_-({\bf r}_\perp)}\left[1-\langle g^{(i)}({\bf r}_\perp)\rangle\right].\label{eq:frac_noise}
\end{align}
In what follows, we describe the AI and PCI techniques, and then derive the ORDI and MORD methods.  For each technique, we compute the expected uncertainty in optical depth resulting from photon shot noise, and conclude with a comparison of these uncertainties across all cases.

\subsection{Absorption imaging} \label{Section:Abs}
AI proceeds directly from the intensities recorded by a single camera placed in-focus (effectively recording the intensity as it was in the object plane at $z=0$), to give the optical depth ${\rm OD}({\bf r}_\perp) = -\ln[1-g({\bf r}_\perp)]$ without further approximation.  AI is most straightforward for resonant ($\bar\delta = 0$) low intensity ($\bar I_-\ll 1$) imaging, giving an optical depth ${\rm OD}({\bf r}_\perp) = \sigma_0 \rho_{\rm 2D}({\bf r}_\perp)$ proportional to the column density.  In this limit, the repeated scattering of resonant photons strongly perturbs the atomic ensemble after a single instance of imaging.  As a result, AI is not generally used to perform weakly destructive measurements\footnote{Reference~\cite{AnandPTAI} describes a weakly destructive absorption technique called partial transfer absorption imaging (PTAI) where only a small fraction of a large atomic ensemble is transferred to a internal state sensitive to a probe laser, while the majority of the atoms were kept in a dark state.}. 

Using the formalism in eqn.~\eqref{eq:frac_noise} along with eqn.~\eqref{eq:OD_Def},  we find the noise for AI to be
\begin{align}
\langle\delta {\rm OD}({\bf r}_\perp)\rangle &= 0, & {\rm and} && \langle\delta {\rm OD}({\bf r}_\perp)\delta {\rm OD}({\bf r}_\perp^\prime)\rangle &= \delta_{{\bf r}_\perp,{\bf r}_\perp^\prime} \frac{\langle\delta g({\bf r}_\perp)^2\rangle}{[1-\langle g({\bf r}_\perp)\rangle]^2}.
\label{AbsImgSqNoise}
\end{align}
This shows that the noise is still $\delta$-correlated in space and diverges at large optical depth.
This is expected because at large OD the vast majority of the probe is absorbed, leading to an increased fractional contribution of photon shot noise. 

\subsection{Phase contrast imaging} \label{Section:PCI}
PCI is often used for weakly destructive measurements, and readily allows the same cloud to be imaged repeatedly. Like AI, PCI is performed with a single in-focus camera, but unlike AI, the recorded intensity of a phase-contrast image contains phase information from which the column density is extracted.  After traversing the layer of atoms, the electric field $E_+ = E_- + \Delta E$ may be expressed as a sum~\cite{Ketterle:ImgNotes} of its unscattered and scattered parts $E_-$ and $\Delta E$.  PCI is typically implemented by adding a $\theta=\pi/2$ phase shift to the unscattered field\footnote{In practice this phase shift is created using a phase plate, with a retarding spot slightly larger than the focused beam spot-size, in the imaging system (Fig.~\ref{Schematic_TwoLensImg}).  The $\pi/2$ phase shift maximizes the phase contrast signal (in contrast, for $\theta$\,=\,$\pi$, the Taylor expansion is dominated by even functions, so that $I \propto \phi^{2}$).} so that $E_- \rightarrow E_- \exp ( i \theta )$.
PCI is typically applied in the far detuned limit where absorption can be neglected, i.e., $\alpha\ll\phi$, giving the PCI intensity pattern
\begin{align}
g({\bf r}_\perp) &= 2 \left[\cos \phi({\bf r}_\perp) - \sin \phi({\bf r}_\perp) - 1\right] \approx -2\phi({\bf r}_\perp) = 2\bar\delta {\rm OD}({\bf r}_\perp) \label{eq:PCApproximate}
\end{align}
resulting from the interference of the the phase-shifted light with the light refracted by the atoms.  The final approximation in Eq.~\eqref{eq:PCApproximate} is valid for small phase shifts (requiring a combination of large detuning), and in this limit the PCI signal is linear in $\phi$, therefore proportional to the optical depth~\cite{Ketterle:97}.

Next, we find the uncertainty in the recovered optical depth for PCI using eqn.~\eqref{eq:PCApproximate} to be
\begin{align}
\langle\delta {\rm OD}({\bf r}_\perp)\rangle &= 0, & {\rm and} && \langle\delta {\rm OD}({\bf r}_\perp)\delta {\rm OD}({\bf r}_\perp^\prime)\rangle &= \delta_{{\bf r}_\perp,{\bf r}_\perp^\prime} \frac{1}{4\bar\delta^2}\langle\delta g({\bf r}_\perp)^2\rangle. \label{PCIsqNoise}
\end{align}
This shows that for large detuning (where the approximations leading to PCI are valid) and low optical depth ($g \ll 1$), the variance in phase contrast imaging is reduced by a factor of $\bar\delta^2$ compared to that of AI.

\subsection{Off resonant defocus imaging} \label{Section:OneCamera}
The AI technique introduced in Sec.~\ref{Section:Abs} is best implemented on resonance, where there are no phase shifts. On the other hand, PCI works best off-resonance when absorption is minimal. The ORDI technique~\cite{Turner:05,Turner:04,Turner:Thesis,Wigley2016a} is a method that works best in the intermediate regime where both absorption and phase shift are important.  ORDI and MORD build from an invertible relation between the observed intensity and the optical depth that recovers all but a small range of spatial frequencies.

ORDI relies on several simplifying assumptions, the first of which is the paraxial approximation to the electric field that has propagated through an atomic cloud in Sec.~\ref{sec:HelmholtzEqn}.  We assume that the electric field did not diffract as it traveled through the cloud (i.e., that it was thin compared to the depth of field).  Both AI and PCI require these same approximations.

Going forward we introduce the Fourier transform (FT) of a two-dimensional function $f({\bf r}_\perp)$ as $\tilde{f}({\bf k}_\perp)$\,=\,$\int_{-\infty}^{\infty} f({\bf r}_\perp) \exp(- i {\bf k}_\perp \cdot {\bf r}_\perp) {\rm d}^2 {{\bf r}_\perp}$.

Using the paraxial propagator and the convolution theorem, we readily obtain the Fresnel diffraction integral
\begin{align}
E({\bf r}_\perp,z) &= \frac{i}{\lambda z} \int_{-\infty}^{\infty} E({\bf r}_\perp,z=0)  \exp \left[ \frac{i k_{0}}{2 z} \left|{\bf r}_\perp - {\bf r}_\perp^\prime\right|^{2} \right] {\rm d}^2 {\bf r}_\perp^\prime. \label{eq:FresnelIntegral}
\end{align}
Using eqn.~\eqref{eq:FresnelIntegral} and the electric field just after it traversed the cloud in eqn.~\eqref{eq:ORDwave}, the normalized intensity at a detector placed an arbitary distance from the atomic ensemble is
\begin{align}
\tilde g ({\bf k}_\perp,z) =& \int_{-\infty}^{\infty} \bigg\{1-\exp [-\alpha ({\bf r}_\perp + z {\bf k}_\perp / 2 k_{0}) -\alpha ({\bf r}_\perp - z {\bf k}_\perp / 2 k_{0})\nonumber\\ &+ i \phi ({\bf r}_\perp -  z {\bf k}_\perp / 2 k_{0}) - i \phi ({\bf r}_\perp +  z {\bf k}_\perp / 2 k_{0}) ]\bigg\} \times \exp(- i {\bf r}_\perp\cdot{\bf k}_\perp) {\rm d}^2 {\bf r}_\perp. 
\end{align}
To derive the ORDI technique we further required that the phase is slowly-varying: $|\phi ({\bf r}_\perp+{ z {\bf k}_\perp / 2 k_{0}}) - \phi ({\bf r}_\perp-{ z {\bf k}_\perp / 2 k_{0}})|  \ll 1$, and that absorption is small: $\alpha ({\bf r}_\perp) \ll 1$. To the lowest order in $\phi$ and $\alpha$, the resulting normalized intensity
\begin{align}
\tilde g ({\bf k}_\perp,z) \approx& \int_{-\infty}^{\infty} [\alpha ({\bf r}_\perp +  z {\bf k}_\perp / 2 k_{0}) +\alpha ({\bf r}_\perp -  z {\bf k}_\perp / 2 k_{0})\\ &- i \phi ({\bf r}_\perp -  z {\bf k}_\perp / 2 k_{0}) + i \phi ({\bf r}_\perp +  z {\bf k}_\perp / 2 k_{0}) ] \times \exp(- i {\bf r}_\perp\cdot{\bf k}_\perp) {\rm d}^2 {\bf r}_\perp\: , \nonumber
\end{align}
is a Fourier integral of $\phi$ and $\alpha$. Defining the FTs of $\phi$ and $\alpha$ as $\tilde{\phi}$ and $\tilde{\alpha}$, respectively, we find
\begin{align}
\tilde g ({\bf k}_\perp,z) &=  2\tilde{\alpha}({\bf k}_\perp) \cos(z k_\perp^2 / 2 k_{0}) - 2\tilde{\phi}({\bf k}_\perp) \sin(z k_\perp^2 / 2 k_{0}),\label{eq:CTF}
\end{align}
where $k_\perp^2 = k_{x}^{2}+k_{y}^{2}$.  Using eq.~\eqref{eq:abs_phase} we arrive at the explicit expression
\begin{align}
\tilde g ({\bf k}_\perp,z)  &= [\cos(z k_\perp^2 / 2 k_{0}) + 2\bar\delta \sin(z k_\perp^2 / 2 k_{0})] \tilde{\rm OD}({\bf k}_\perp) \equiv \tilde{h}({\bf k}_\perp, z)  \tilde{\rm OD}({\bf k}_\perp),\label{eq:ORDIresult}
\end{align}
that uses a CTF $\tilde{h}({\bf k}_\perp, z)$ to provide a linear relation between the Fourier transformed optical depth $ \tilde{\rm OD}({\bf k}_\perp)$ and the fractional change in intensity $\tilde g ({\bf k}_\perp,z)$ defined in eq.~\eqref{eq:FracIntensity}.

We plot a representative CTF and its inverse in Fig.~\ref{CTFOneFullFig} showing singularities at some spatial frequencies $k_\perp$, the locations of which depend upon the camera position $z$ and the detuning $\bar\delta$. When $\tilde{h}({\bf k}_\perp, z)^{-1}$ is applied, measurement noise is amplified near the divergences, so that no useful information can be extracted from these spatial frequencies.  The ratio $\alpha / \phi = -1/2\bar\delta$ along with the sign of $z$ determines the quality of information at low spatial frequencies. When ${\rm sign}(z) \alpha / \phi < 0$, the inverse CTF has a divergence at a low spatial frequency, and information may be lost for the spatial structure of interest.  There are two mathematically equivalent cases to achieve the ``good'' condition: $\bar\delta < 0$ (red detuning) with $z < 0$ (negative defocus), and $\bar\delta > 0$ (blue detuning) with $z > 0$ (positive defocus). 

\subsubsection{ORDI statistical uncertainties and regularization} \label{sec:reg}

Here we compute the anticipated diverging uncertainty at the zeros of the CTF quantitatively and introduce a regularization parameter to resolve these divergences.  In general the spatial structure in eq.~\eqref{eq:frac_noise} would lead to correlated noise in $\langle\delta \tilde g({\bf k}_\perp)\delta \tilde g({\bf k}_\perp^\prime)\rangle$, however, for simplicity we assume that the spatial dependence is weak (as would be the case for an extended system).  We are reminded that random variables that are uniform and $\delta$-correlated spatially are also uniform and $\delta$-correlated in the Fourier basis, giving $\langle\delta \tilde g({\bf k}_\perp)\delta \tilde g({\bf k}_\perp^\prime)\rangle = \delta_{{\bf k}_\perp,{\bf k}_\perp^\prime}\langle\delta \tilde g({\bf k}_\perp)^2\rangle$.

The anticipated variance of the reconstructed optical depth is therefore
\begin{align*}
\langle\delta {\rm OD}({\bf k}_\perp)\delta {\rm OD}({\bf k}_\perp^\prime)\rangle &= \delta_{{\bf k}_\perp,{\bf k}_\perp^\prime}\tilde{h}({\bf k}_\perp, z)^{-2}\langle\delta \tilde g({\bf k}_\perp)^2\rangle,
\end{align*}
divergent at the zeros of $\tilde{h}({\bf k}_\perp, z)$.

Here we describe the process of regularization used to mitigate the amplification of noise near divergences in the inverse CTF.  The basic idea is to include a ``default'' (i.e., a Bayesian prior) optical depth ${\rm OD}_0({\bf k}_\perp)$ with uncertainty quantified by a known $\delta$-correlated random variable $\delta {\rm OD}_0({\bf k}_\perp)$, and we use the weighted average of the reconstructed and prior images
\begin{align}
\tilde{\rm OD}({\bf k}_\perp) &= \frac{a({\bf k}_\perp) \tilde{h}({\bf k}_\perp, z)^{-1} \tilde g ({\bf k}_\perp,z) + b({\bf k}_\perp) {\rm OD}_0({\bf k}_\perp)}{a({\bf k}_\perp) + b({\bf k}_\perp)}
\end{align}
as our reconstructed optical depth.
We determine the coefficients $a({\bf k}_\perp)$ and $b({\bf k}_\perp)$ by minimizing the variance 
\begin{align*}
\langle\delta {\rm OD}({\bf k}_\perp)\delta {\rm OD}({\bf k}_\perp^\prime)\rangle &= \delta_{{\bf k}_\perp,{\bf k}_\perp^\prime}\frac{a({\bf k}_\perp)^2\tilde{h}({\bf k}_\perp, z)^{-2}\langle(\delta \tilde g_\perp)^2\rangle + b({\bf k}_\perp)^2 \langle\delta {\rm OD}_0({\bf k}_\perp)^2\rangle}{[a({\bf k}_\perp) + b({\bf k}_\perp)]^2},
\end{align*}
for each wavevector.
This gives standard expression for Wiener deconvolution~\cite{Orieux2010}
\begin{align}
\tilde{\rm OD}({\bf k}_\perp) &= \frac{\tilde{h}({\bf k}_\perp, z) \tilde g ({\bf k}_\perp,z) + \eta^2({\bf k}_\perp) {\rm OD}_0({\bf k}_\perp)}{\tilde{h}({\bf k}_\perp, z)^2 + \eta({\bf k}_\perp)^2}
\end{align}
with regularization constant $\eta({\bf k}_\perp)^2 = \langle\delta \tilde g({\bf k}_\perp)^2\rangle / \langle\delta {\rm OD}_0({\bf k}_\perp)^2\rangle$.
In practice we typically select the prior ${\rm OD}_0({\bf k}_\perp) = 0$, essentially assuming no prior information, and take $\eta$ to be ${\bf k}$-independent.
This results in the Tikhonov regularized inverse CTF
\begin{align}
\tilde{h}_{\rm R}({\bf k}_\perp, z)^{-1} &= \frac{\tilde{h}({\bf k}_\perp, z)}{\tilde{h}({\bf k}_\perp, z)^2 + \eta^2} \label{eq:CTFinv}
\end{align}
that blocks the divergence of statistical noise, but introduces artifacts.  

Since the ORDI technique is implemented in the Fourier basis, we find the noise-correlation function in coordinate space
\begin{align}
\langle\delta {\rm OD}({\bf r}_\perp)\delta {\rm OD}({\bf r}_\perp^\prime)\rangle &= \langle\delta g({\bf r}_\perp)^2\rangle \frac{1}{N}\sum_{{\bf k}_\perp} \frac{\tilde{h}({\bf k}_\perp, z)^2}{\left[\tilde{h}({\bf k}_\perp, z)^2 + \eta^2\right]^2} e^{i {\bf k}_\perp\cdot\left({\bf r}_\perp-{\bf r}_\perp^\prime\right)} \label{MORDsqNoise}
\end{align}
where we explicitly expressed the discrete FT in terms of a sum over a total of $N$ pixels.  Unlike AI or PCI, ORDI reconstruction of the optical depth has correlated noise.

\subsection{Multi-camera off resonant defocus imaging} \label{Section:ThreeCamera}

\begin{figure}[t!]
\begin{center}
\includegraphics{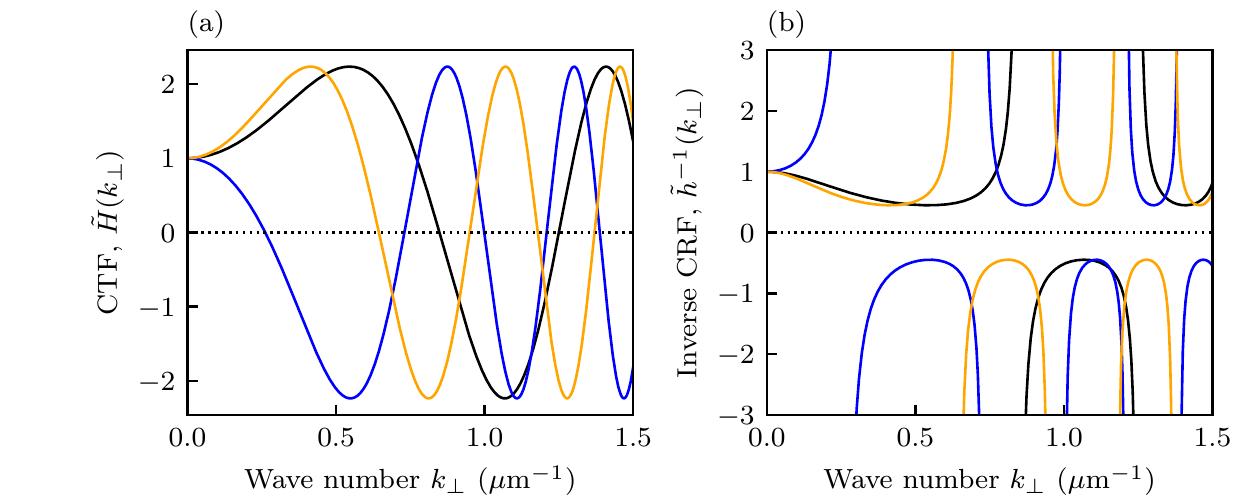}
\end{center}
\caption[Multiple contrast transfer functions]{Forward (a) and inverse (b) CTFs computed as a function of transverse wave number $k_\perp$, for detuning $\bar\delta = 1$ for three defocus distances $60\ \mu{\rm m}$ (black), $-100\ \mu{\rm m}$ (blue), and $87\ \mu{\rm m}$ (orange).
This shows that for a generic selection of defocus distances the CTFs do not have coincident zeros}
\label{CTFThreeFullFig}
\end{figure}

We showed in Sec.~\ref{Section:OneCamera} that with the ORDI method, spatial frequencies exist for which we retrieve no information (where $\tilde{h} \rightarrow 0$). These spatial frequencies depend on $z$, the position in the object plane where the intensities are recorded (Fig.~\ref{CTFThreeFullFig}). This suggests that by adding cameras to the system at different $z$, we might recover information at all spatial frequencies.
In the multiple camera method, the images are combined and processed in a way that minimizes the uncertainty in the recovered optical depth.

As shown in Sec.~\ref{Section:OneCamera}, each CTF taken independently provides a linear relation between a measurement and the estimated optical depth [eqn.~\eqref{eq:ORDIresult}]. Our task is to find the best estimate of the optical depth for MORD using the measurements taken in the laboratory [eqn.~\eqref{eq:FracIntensity}] and their corresponding CTFs. We model shot-noise, the only source of noise we consider, as noise that is spatially uncorrelated and uniform over all spatial frequencies (i.e., additive white noise).  Following the argument in Sec.~\ref{sec:reg}, we minimize the noise in the reconstructed optical depth and find
\begin{equation}
\tilde{\rm OD}_{k} = -\frac{\sum_{(i)} \tilde{h}_{k}^{(i)} \tilde{g}_{k}^{(i)}} { \sum_{(i)} |\tilde{h}_{k}^{(i)}|^{2} + \eta^{2} }.\label{eq:ThreeCameraReconstruction}
\end{equation}
For $\eta=0$ and non-overlapping divergences, the above equation is an exact solution to the intensity predicted by the scalar wave equation, subject to the approximations discussed above.  As with the ORDI case, this transfer function also transforms uncorrelated measurement noise into correlated noise in reconstructed images.

\subsection{Comparison of statistical and systematic uncertainties} \label{Sec:AllImgMethods}

\begin{figure}[t!]
\begin{center}
\includegraphics{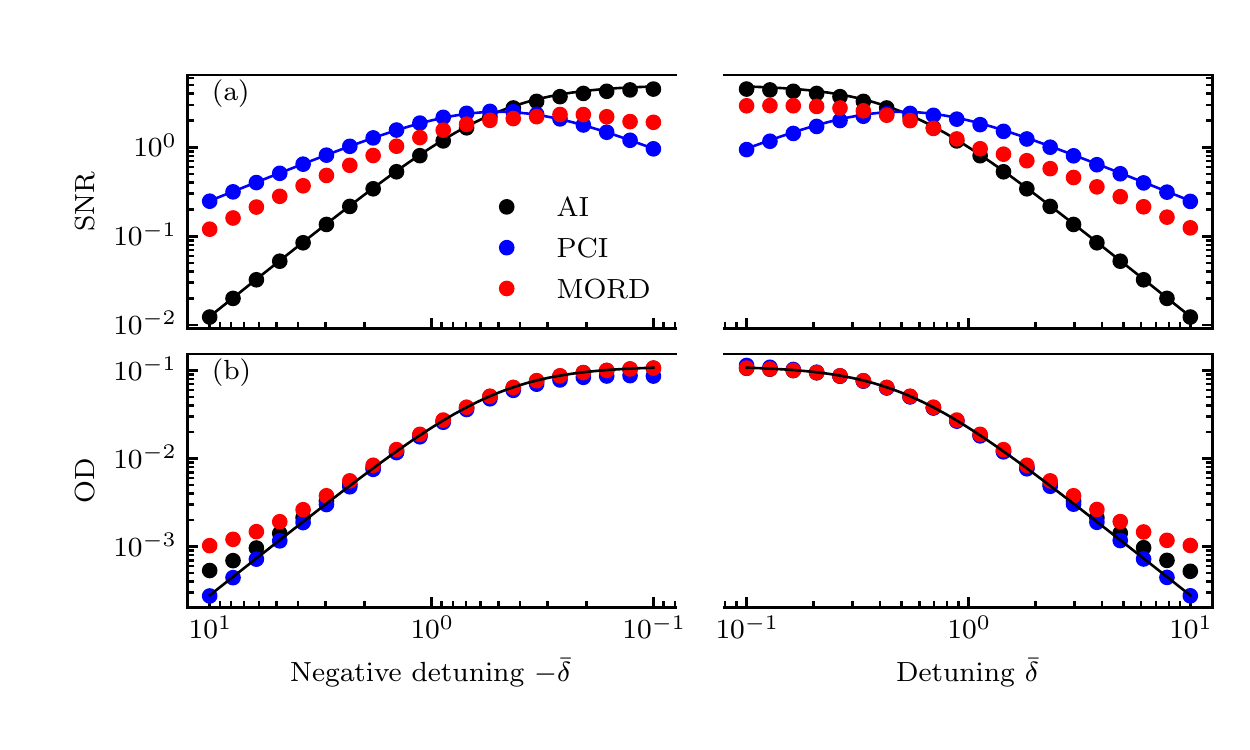}
\end{center}
\caption[ Comparing different imaging techniques]{Simulations.  (a) Computed signal to noise ratio for AI (black markers), PCI (blue markers), and MORD (red markers)  for both red and blue detunings plotted on separate log scales. 
The numerical simulations and analytical solutions (see Sects.~\ref{Section:Abs} and \ref{Section:PCI}) are shown as symbols and solid curves, respectively.  These simulations combined three defocus distances $60\ \mu{\rm m}$, $-100\ \mu{\rm m}$, and $87\ \mu{\rm m}$ and set the regularization parameter $\eta = 0$.
(b) Optical depth obtained for AI, PCI, and MORD compared to the true value (black curve).
Both MORD and AI show a systematic shift at small OD resulting from noise rectification when converting from measured intensities to OD.}
\label{fig:UncertaintiesAllMethods}
\end{figure}

We compare the signal to noise ratio (SNR) of three-camera MORD to those of AI and PCI in Fig.~\ref{fig:UncertaintiesAllMethods}(a). 
The solid symbols are the result of a numerical simulation of our full imaging process, while the curves for PCI and AI plot the expected SNR given the expressions in eqns.~\eqref{AbsImgSqNoise} and \eqref{PCIsqNoise}, 
thereby confirming the performance of our numerical model.
Except for the resonant case, the MORD imaging technique gives a larger SNR than AI at the same detuning.
Further, for positive detuning [right panel to Fig.~\ref{fig:UncertaintiesAllMethods}(a)] MORD method shows the same $\delta^{-1}$ scaling in the SNR as PCI, but with about $2\times$ lower SNR.
The performance at negative detuning is slightly reduced owing to two of the cameras having a near-zero $k$ node in their CTF as compared to just one for positive detuning.

Next Fig.~\ref{fig:UncertaintiesAllMethods}(b) quantifies systematic uncertainties in the MORD calculated optical depth.
Here the black curve plots the true optical depth, and the symbols plot the outcome of our numerical simulation.
At very low optical depth both AI and MORD begin to diverge from the true signal.
These systematic shifts result from the non-linearity in the logarithm used to convert from fractional absorption to OD, where detector noise is rectified, leading to excess signal.

\section{Experimental techniques} \label{Sec:Techniques}

\begin{figure}[t!]
\begin{center}
\includegraphics{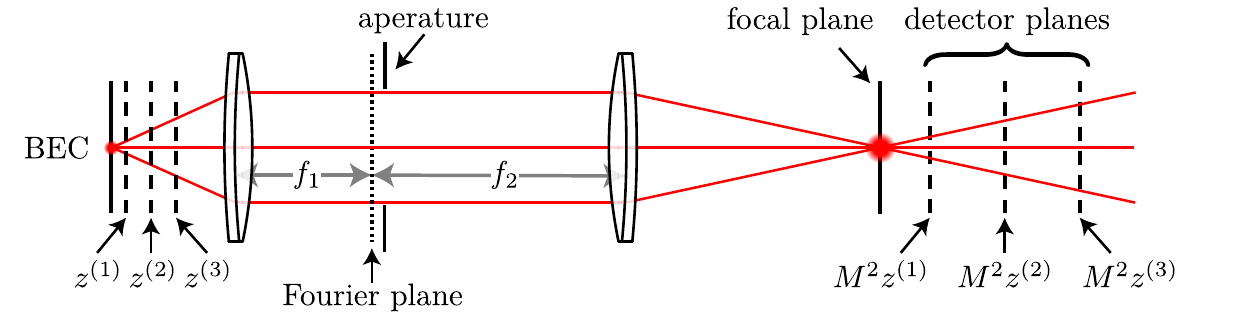}
\end{center}
\caption[Schematic of a two-lens Keplerian imaging system]{A simplified schematic of a two-lens Keplerian imaging system with magnification $M\approx3.3$ using lenses $f_1 = 75\ {\rm mm}$ and $f_2 = 250\ {\rm mm}$. 
In MORD, the three detector planes $z^{(i)}$ are located away from the focus.
This is equivalent to having three detectors next to BEC, as shown.
The imaging resolution is set by the diameter of the apertures placed in the imaging system.
A representative aperture, placed near the Fourier plane, is shown above.}
\label{Schematic_TwoLensImg}
\end{figure}

Our optical geometry, schematically depicted in Fig.~\ref{Schematic_TwoLensImg}, consisted of a standard two-lens Keplerian microscope.
The first lens (focal length $f_{1}$) was positioned a distance $f_{1}$ from the object (a BEC in our experiments); and the second lens (focal length $f_{2}$) was placed a distance $f_{1} + f_{2}$ from the first lens. 
Typically the detector would be placed at the focus of the second lens, and would have a magnification $M = f_{2}/f_{1}$.
This system forms the conceptual basis of the imaging system described in this manuscript.

\begin{figure}[t!]
\begin{center}
\includegraphics{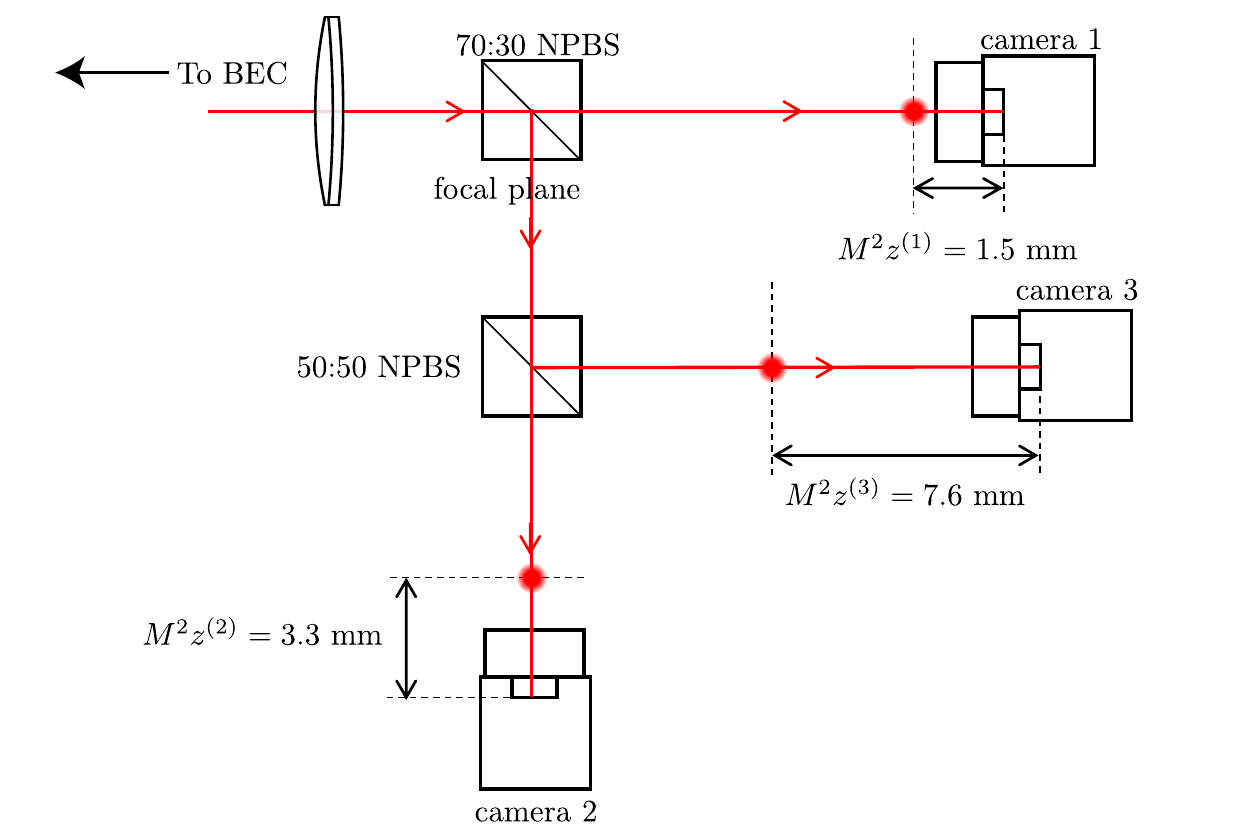}
\end{center}
\caption[Schematic of the MORD technique]{Schematic of our MORD setup. The two-lens imaging system described by Fig.~\ref{Schematic_TwoLensImg} is followed by two non-polarizing beam splitters that nominally split the probe laser beam into three equal intensities.
Each image is then recorded by its designated camera.
Each camera takes a defocused image at position $M^{2} z^{(i)}$ from focus, where the red circles denote the image planes.}
\label{fig:Schematic_ExperimentSetUp}
\end{figure}

We implemented the MORD technique using a two-lens imaging system with magnification $M\approx3.3$ (as shown in Fig.~\ref{Schematic_TwoLensImg}), and placed three detectors at different distances from the focal plane.
The system consisted of a pair of 25.4~mm diameter lenses with focal lengths $f_{1} = 75$~mm and $f_{2} = 250$~mm separated by 325(10)~mm.
In such an imaging system, a point $(x_{0},y_{0},z_{0})$ in the object space is imaged to the point $(-M x_{0},-M y_{0},M^{2} z_0)$ in the image space.
We placed a $18(1)$~mm aperture close to the imaging system's Fourier plane to minimize spherical aberrations.
This gave a numerical aperture NA $\approx0.12$.
After the final lens, roughly equal fractions of the light was directed to each of our three detectors~(Fig.~\ref{fig:Schematic_ExperimentSetUp}) using non-polarizing beam splitters (NPBSs) with reflection to transmission (R:T) ratios 70:30 and 50:50.
We detected this light on charge-coupled device (CCD) cameras with 648$\times$488 square pixels with width 5.6~$\mu$m.
Each camera was on a translation stage, so that the set-up could be used for both the defocused and standard absorption imaging methods.

\subsection{Experimental procedure} \label{Section:Procedure}

Here we describe the experimental procedure used to acquire the images for the MORD technique.
We collected about $2\times10^8$ $^{87}$Rb atoms in a vapor-fed six-beam magneto-optical trap, performed sub-Doppler cooling, and then trapped the atoms in the $\left|f = 2, m_{F} = 2\right\rangle$ state in a spherical quadrupole trap.
We used magnetic transport~~\cite{Greiner:Transport} to move the resulting cloud about 42~cm vertically in 2.2~s, giving an ensemble at $\approx 120\ \mu$K with about $1\times10^{8}$ atoms.
We then evaporated to degeneracy in the combined magnetic/optical technique described in Ref.~\cite{Lin:PRA}.
During evaporation, we performed a microwave transfer between the ground hyperfine states $\left|f = 2, m_{F} = 2\right\rangle$ to $\left|f = 1, m_{F} = -1\right\rangle$, giving $100\times10^{3}$ atom BECs in a cross-optical dipole trap every $15$~s.
The resulting trap frequencies were $\approx$ 110~Hz, 75~Hz, and 50~Hz along $\ex$,$\ey$, and $\ez$ respectively.
To achieve the desired optical depth to test our imaging technique, we performed a partial $\approx10$~\% transfer to the $\left|f = 2, m_{F} = 1\right\rangle$ state with a short ($\ll \pi /2$) microwave pulse~\cite{AnandPTAI}.
We then used a probe beam with $\lambda\approx780\ {\rm nm}$ detuned a variable $\bar\delta$ from the $f$\,=\,$2$ to $f'$\,=\,$3$ cycling transition (without a repumping laser) to image the transferred $N\approx1\times10^4$ atoms after a 6~ms time of flight (TOF) with the three cameras simultaneously.
In Sec.~\ref{subsec:Data}, we present the results of this experiment for $\bar\delta = 2.0(1)$.
We used a probe beam with intensity $I\approx 2.5 I_{\rm sat}$, where $I_{\rm sat}\approx1.67~{\rm mW/cm^{2}}$~\cite{steck2001rubidium} for a circularly polarized probe beam.

\section{Measurement and analysis} \label{sec:MeasAnalysis}

\subsection{Experimental data} \label{subsec:Data}
\begin{figure}[t!]
\begin{center}
\includegraphics{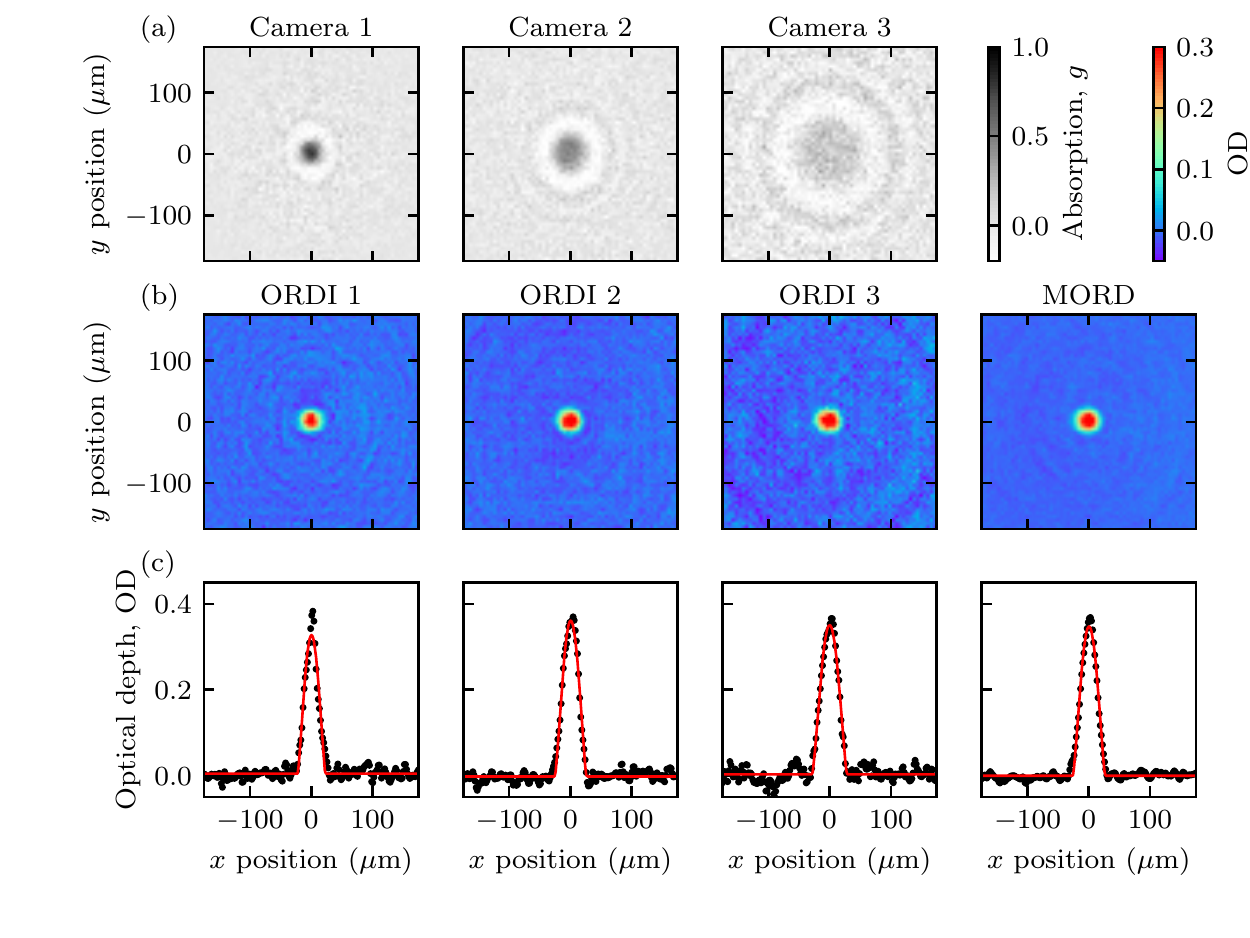}
\end{center}
\caption[Raw and reconstructed data]{Raw and refocused data.
(a) Raw data with defocus distances of $M^2 z^{(1)} = 1.5\ {\rm mm}$, $M^2 z^{(2)} = 3.3\ {\rm mm}$ and $M^2 z^{(3)} = 7.6\ {\rm mm}$ showing clear defocus patterns.
A radial mask excluded artifacts from the probe beam diffracting off of the edge of the CCD sensor.
(b) Refocused images.  
The first three images depict individual ORDI reconstructions and the right-most image shows a MORD reconstruction.
The ring-like artifacts present in the ORDI method are less prevalent in the MORD data.
(c) Cross sections along $\ex$ (black markers) along with fits (red curves) to the expected ``Thomas-Fermi'' distribution.
}
\label{fig:Data1}
\end{figure}

Fig.~\ref{fig:Data1}(a) shows the fractional absorption $g^{(i)}$ recorded on each camera as described by eqn.~\eqref{eq:FracIntensity}.
In practice we obtained a third image with neither the atoms nor the probe light.
This background image was subtracted from $I^{(i)}_+$ and $I^{(i)}_-$ to remove dark counts and stray light from the images.
The camera 1 image was defocused a distance $z^{(1)}$\,=\,$1.5$\,mm away from the image plane.
Similarly, the diffraction patterns from cameras 2 and 3  resulted from defocus distances of $z^{(2)}$\,=\,$3.3$\,mm and $z^{(3)}$\,=\,$7.6$\,mm, respectively.
These distances were chosen to insure the diffraction patterns were discernible within the imaging system resolution, and the interference pattern was contained within the spatial extent of the detector.

To compare the ORDI and MORD imaging techniques, we show the single-camera ORDI results for each of the three cameras independently in the left three panels of Fig.~\ref{fig:Data1}(b), and the three-camera MORD reconstructed on the right panel.
The background of the MORD method has visibly fewer artifacts than the ORDI reconstructed images.
Lastly, we address the fidelity of the reconstruction by comparing to the expected ``Thomas-Fermi'' density distribution~\cite{Castin1996}
\begin{align}
n(x) = n_0 \left[1 - (x-x_0)^2/r_{\rm TF}^2\right]^{3/2}.
\end{align}
Figure \ref{fig:Data1}(c) plots cross-sectional cuts along $\ex$ through the reconstructions with black symbols along with fits to the Tomas-Fermi model with red curves.
We see that the data is in good agreement with the expected behavior.

The defocus distances $z^{(i)}$ used in the ORDI and MORD reconstructions shown in Fig.~\ref{fig:Data1} were first measured in the lab with $\approx 1\ {\rm mm}$ uncertainties, we then used the focusing technique in Ref.~\cite{Putra:14} to fine tune the displacements $z^{(i)}$.

\section{Conclusion}
We experimentally demonstrated an improvement to the single-camera ORDI technique by using three cameras simultaneously to eliminate the divergences that arise in any single CTF. We studied the systematic uncertainties of the MORD method, and theoretically compared multiple techniques using simulated data to understand the array of imaging techniques on equal footing. We showed that in the regime of far detuned probe beam and low optical depth, the MORD method is comparable to PCI. Therefore, in experiment, the easier to implement MORD set-up may be preferable to PCI.
 
Because the first low-spatial-frequency divergence of a CTF may occur at length scales of tens of micrometers, ORDI has been limited to imaging relatively large objects.
By eliminating the spatial CTF divergences, MORD offers a complementary technique to in-situ AI and PCI imaging of quantum gases.
The SNR of MORD imaging is about 1.5 times below both AI and PCI in their respective optimal operating conditions: resonant imaging for AI and far-detuned imaging for PCI.
As a result MORD can switch between near optimal strong (destructive) measurements and weak (minimally destructive) measurements with no hardware modifications.
In addition, MORD eliminates the need to refocus the imaging systems, for example in TOF experiments where the object plane shifts depending on the TOF.

A final noteworthy outcome of our noise analysis is that even with spatially uncorrelated detector noise, the ORDI and MORD methods introduce correlations into the reconstructed signal, meaning that without added calibration, these techniques would not be of use in experiments studying noise-correlations between atoms~\cite{ChengChing:NJP2011}.
\newline
\newline
\textbf{Funding}
This work was partially supported by the AROs Atomtronics MURI, by the AFOSRs Quantum Matter MURI, NIST, and the NSF through the PFC at the JQI.
\newline
\newline
\textbf{Acknowledgments}
We thank D. Barker and A. Putra for their careful and thorough reading of this manuscript.
We appreciate the efforts of E.~Altuntas, R.~P.~Anderson, Q.-Y. Liang, D.~Trypogeorgos, and A.~Valdés-Curiel for employing defocus imaging in their experiments, and motivating the completion of this manuscript.
\newline
\newline
\textbf{Disclosures}
The authors declare no conflicts of interest.
\newline
\newline
\textbf{Data Availability Statement}
Data underlying the results presented in this paper are not publicly available at this time but may be obtained from the authors upon reasonable request.

\bibliography{Bibliography}

\end{document}